\documentclass[aps,twocolumn,prb,showpacs]{revtex4}
\usepackage{graphicx}
\begin{document}
\draft
\preprint{14 January 2006}
\title{Evidence for Multimagnon-Mediated Nuclear Spin Relaxation in the
       Intertwining Double-Chain Ferrimagnet Ca$_3$Cu$_3$(PO$_4$)$_4$}
\author{Shoji Yamamoto, Hiromitsu Hori, Yuji Furukawa, Yusuke Nishisaka,
        Yuzuru Sumida, Kiyotaka Yamada, and Ken-ichi Kumagai}
\address{Division of Physics, Hokkaido University,
         Sapporo 060-0810, Japan}
\author{Takayuki Asano}
\address{Department of Physics, Kyushu University,
         Fukuoka 812-8581, Japan}
\author{Yuji Inagaki}
\address{Department of Chemistry, Kyushu University,
         Fukuoka 812-8581, Japan}
\date{\today}
\begin{abstract}
The nuclear spin-lattice relaxation time $T_1$ of $^{31}\mbox{P}$ nuclei
in the title compound is measured for the first time and interpreted in
terms of a modified spin-wave theory.
We establish a novel scenario for one-dimensional ferrimagnetic spin
dynamics---it is {\it three-magnon processes enhanced by exchange
scattering}, rather than Raman processes, that make the major contribution
to $1/T_1$.
\end{abstract}
\pacs{76.50.$+$g, 76.60.$-$k, 75.50.Gg}
\maketitle

\section{Introduction}

   Low-energy dynamics in one-dimensional (1D) quantum magnets is
a long-standing problem and nuclear magnetic resonance (NMR) findings
are often enlightening in this context.
It was a fine collaboration that the NMR relaxation rates
$1/T_1$ and $1/T_{2{\rm G}}$ (Ref. \onlinecite{T13618}) of
the spin-$\frac{1}{2}$ antiferromagnet Sr$_2$CuO$_3$ illuminated
multiplicative logarithmic corrections to the dynamic susceptibility
\cite{S539} in critical spin chains.
$T_1$ measurements on the spin-$\frac{5}{2}$ antiferromagnet
(CH$_3$)$_4$NMnCl$_3$ were a pioneering study in an attempt to detect the
long-time diffusive spin dynamics in one dimension. \cite{B2215}
The diffusive contribution to $1/T_1$ was observed in the
spin-$\frac{1}{2}$ quantum limit \cite{T4612,K2765} and for spin-gapped
antiferromagnets \cite{T2173} as well.

   Antiferromagnets, whether critical or gapped, are thus vigorously
studied, while little is known about ferrimagnetic, as well as
ferromagnetic, dynamics.
Recently numerous 1D ferrimagnets have been synthesized in an effort to
design molecule-based ferromagnets.
The static properties of various heterospin chains were correspondingly
calculated, \cite{D10992,K3336,T5355,B3921,P8894,Y14008,I14024,Y1024} but
few effort has been devoted to exploring their dynamic features.
Recent NMR observations of 1D ferrimagnets \cite{F8410,F433,F064422} have
thus been motivated and have indeed stimulated the theoretical interest in
them.
The most pioneering $T_1$ measurements, \cite{F8410} performed on
the metal-radical hybrid compound
Mn(C$_5$H$_2$O$_2$F$_6$)$_2$C$_{10}$H$_{17}$N$_2$O$_2$, were interpreted
in terms of solitonic excitations at low temperatures and based on the
spin-diffusion model at high temperatures.
However, this material has rather large exchange interactions and its
magnetic susceptibility ($\chi$) times temperature ($T$) exhibits no
ferrimagnetic minimum at a measurable temperature. \cite{C1756}
The nonnegligible single-ion anisotropy and the spread magnetic moment
over the radical may also blur the intrinsic ferrimagnetic features.
The family of manganeseporphyrin-based ferrimagnets, whose exchange
interactions are somewhat smaller, was also observed through NMR,
\cite{F064422} but the complicated crystalline structure restricted the
analysis to solving the magnetic configuration.
$^1\mbox{H}$ NMR findings \cite{F433} on the bimetallic chain compound
NiCu(C$_7$H$_6$N$_2$O$_6$)(H$_2$O)$_3$$\cdot$2H$_2$O implied a spin-wave
relaxation scenario, but the averaging effect over the numerous protons
masks the low-temperature dynamics.
\begin{figure}
\centering
\includegraphics[width=60mm]{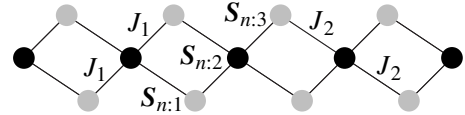}
\vspace*{-3mm}
\caption{Cu$^{\rm II}$ trimeric chains in Ca$_3$Cu$_3$(PO$_4$)$_4$.
         The strongly coupled Cu$^{\rm II}$ trimer consists of a
         central square planar Cu(1) ion (black circle) and two
         pyramidal Cu(2) ions (gray circles).}
\label{F:illust}
\end{figure}
\begin{figure}
\centering
\includegraphics[width=84mm]{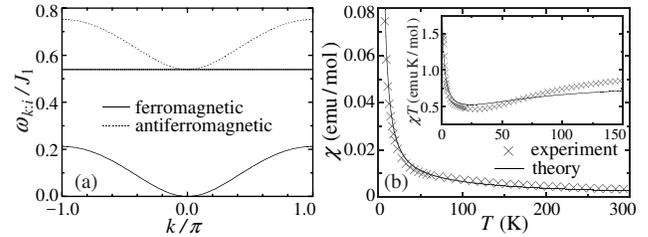}
\vspace*{-3mm}
\caption{(a) Dispersion relations of the spin-wave excitations.
         (b) Modified spin-wave calculations of the susceptibility
             compared with experimental findings at $H=0.1\,\mbox{T}$.}
\label{F:chi}
\end{figure}

   In such circumstances, we perform an NMR study on the homometallic
chain compound Ca$_3$Cu$_3$(PO$_4$)$_4$ (Fig. \ref{F:illust}), which is a
novel ferrimagnet of topological origin. \cite{D83}
This collaboration aims at verifying a newly developed modified spin-wave
nuclear magnetic relaxation theory. \cite{H1453}
Two of the present authors have recently demonstrated that $1/T_1$ in 1D
heterometallic ferrimagnets may significantly be enhanced by exchange
interactions, in an attempt to interpret the novel NMR observations
\cite{F433} for NiCu(C$_7$H$_6$N$_2$O$_6$)(H$_2$O)$_3$$\cdot$2H$_2$O,
which are hardly understandable within the usual Raman relaxation scheme.
However, all the findings \cite{H9023} were far from conclusive to a
possibility of more than two magnons mediating the proton spin relaxation.
Since only the $^1\mbox{H}$ nuclei were practically available as probes in
NiCu(C$_7$H$_6$N$_2$O$_6$)(H$_2$O)$_3$$\cdot$2H$_2$O, the crystal water of
nuisance and the resultant ill-behaving spin-echo recovery curve with
increasing field at low temperatures restricted the $T_1$ analysis to
rather high temperatures $T\agt 60\,\mbox{K}$.
Ca$_3$Cu$_3$(PO$_4$)$_4$ contains P atoms as efficient probes.
The well-isolated Cu$^{\rm II}$ chains with Ca columns in between,
no single-ion anisotropy of the Cu$^{\rm II}$ spins, and the ferrimagnetic
minimum of $\chi T$ at a moderate temperature [see Fig. \ref{F:chi}(b)]
furthermore guarantee this material to be a fine stage of 1D ferrimagnetic
dynamics.

\section{Multimagnon-Mediated Nuclear Spin Relaxation}

   We describe Ca$_3$Cu$_3$(PO$_4$)$_4$ by the Hamiltonian,
\begin{eqnarray}
   &&
   {\cal H}
   =\sum_{n=1}^N
    \bigl[
     J_1(\mbox{\boldmath$S$}_{n:1}\cdot\mbox{\boldmath$S$}_{n:2}
        +\mbox{\boldmath$S$}_{n:2}\cdot\mbox{\boldmath$S$}_{n:3})
   \nonumber\\
   &&\qquad
    +J_2(\mbox{\boldmath$S$}_{n+1:1}\cdot\mbox{\boldmath$S$}_{n:2}
        +\mbox{\boldmath$S$}_{n:2}\cdot\mbox{\boldmath$S$}_{n-1:3})
   \nonumber\\
   &&\qquad
    -g\mu_{\rm B}H(S^z_{n:1}+S^z_{n:2}+S^z_{n:})
    \bigr],
   \label{E:H}
\end{eqnarray}
where each Cu(1) ion is antiferromagnetically coupled to four
Cu(2) ions in an applied field ($H$) (see Fig. \ref{F:illust}).
We set $J_1/k_{\rm B}$ and $J_2/k_{\rm B}$ equal to $100\,\mbox{K}$ and
$8\,\mbox{K}$, respectively. \cite{D83}
Employing the Holstein-Primakoff transformation, \cite{H1098} we expand
the Hamiltonian with respect to $1/S$ as
\begin{equation}
   {\cal H}=-2S^2(J_1+J_2)N+{\cal H}_1+{\cal H}_0+O(S^{-1}),
\end{equation}
where ${\cal H}_i$ contains the $O(S^i)$ terms.
${\cal H}_1$ describes linear spin-wave excitations and is diagonalized
in the momentum space as
\begin{equation}
   {\cal H}_1
   =-\frac{3}{2}(J_1+J_2)
    +\sum_k
     \Biggl(
      \omega_k
     +\sum_{i=1}^3
      \omega_{k:i}\alpha_{k:i}^\dagger\alpha_{k:i}
     \Biggr),
\end{equation}
where $\alpha_{k:i}^\dagger$ creates a spin wave of ferromagnetic
($i=1,2$) or antiferromagnetic ($i=3$) aspect, whose excitation energy is
given by
\begin{equation}
   \left.
    \begin{array}{lll}
     \omega_{k:1}&=&\omega_k-{\displaystyle\frac{S}{2}}(J_1+J_2)
                  + g\mu_{\rm B}H,\\
     \omega_{k:2}&=&(J_1+J_2)S+g\mu_{\rm B}H,\\
     \omega_{k:3}&=&\omega_k+{\displaystyle\frac{S}{2}}(J_1+J_2)
                  - g\mu_{\rm B}H,\\
    \end{array}
   \right.
\end{equation}
with
\begin{equation}
   \omega_k
   =\frac{S}{2}\sqrt{(J_1+J_2)^2+32J_1J_2\sin^2\frac{k}{2}},
\end{equation}
[see Fig. \ref{F:chi}(a)].
${\cal H}_0$ gives two-body interactions and makes a crucial contribution
to nuclear spin-lattice relaxation.
The $O(S^{-1})$ terms are neglected in the following.
The dispersive branches $\omega_{k:1}$ and $\omega_{k:3}$ are
reminiscent of the dual excitations in alternating-spin chains,
\cite{Y13610} whereas the flat band $\omega_{k:2}$, describing
intratrimer excitations, is peculiar to the present system.

   Our way \cite{Y14008,Y11033} of modifying the conventional spin-wave
theory is distinct from the original idea proposed by Takahashi
\cite{T2494} and Hirsch {\it et al.}. \cite{H4769}
Their way of suppressing the divergent sublattice magnetizations consists
of diagonalizing the Hamiltonian together with a Lagrange multiplier
subject to zero staggered magnetization.
The thus-obtained energy spectrum depends on temperature and fails, for
instance, to reproduce the Schottky-peaked specific heat.
Seeking after better thermodynamics, we diagonalize the bare
Hamiltonian and then minimize the free energy with a Lagrange multiplier
subject to zero staggered magnetization. \cite{Y064426}
The thus-calculated $\chi$ is in good agreement with observations
[Fig. \ref{F:chi}(b)].
\begin{figure*}
\centering
\includegraphics[width=160mm]{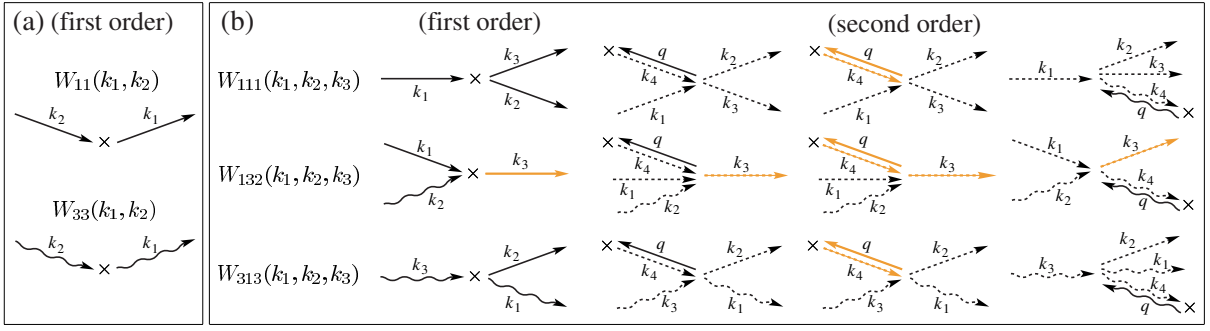}
\vspace*{-3mm}
\caption{(Color online)
         Various nuclear spin-lattice relaxation processes.
         Spin waves which are emitted in the first-order mechanism (solid
         arrows) flip a nuclear spin ($\times$) via the hyperfine
         interaction, whereas four-magnon exchange correlations (dotted
         arrows) thermally scatter a first-order {\it virtual spin wave}
         with $q=-k_4$, where ferromagnetic spin waves are drawn by black
         ($\omega_{k:1}$) and colored ($\omega_{k:2}$) straight arrows,
         while antiferromagnetic ones ($\omega_{k:3}$) by wavy arrows.
         (a) First-order Raman processes.
         (b) First- and second-order three-magnon processes, which are
             inseparable in nonlinear equations.}
\label{F:diagram}
\end{figure*}

   The hyperfine interaction between a $^{31}\mbox{P}$ nucleus and
Cu$^{\rm II}$ spins consists of isotropic Fermi contact and anisotropic
dipolar coupling and is defined as
\begin{equation}
   {\cal H}_{\rm hf}
   =g\mu_{\rm B}\hbar\gamma_{\rm N}I^+
    \sum_n\sum_{i=1}^3
    (\frac{1}{2}A_{n:i}^-S_{n:i}^-+A_{n:i}^zS_{n:i}^z).
\end{equation}
${\cal H}_0$ and ${\cal H}_{\rm hf}$ are both much smaller than
${\cal H}_1$ and may be regarded as perturbations to the linear spin-wave
system.
When we calculate up to second order in
${\cal V}\equiv{\cal H}_0+{\cal H}_{\rm hf}$, the probability of a nuclear
spin being scattered from the state of $I^z=m$ to that of $I^z=m+1$ is
given by
\begin{equation}
   W=\frac{2\pi}{\hbar}\sum_f
     \Biggl|\Bigl\langle f\Bigl|
      {\cal V}+\sum_{m(\neq i)}
      \frac{{\cal V}|m\rangle\langle m|{\cal V}}{E_i-E_m}
     \Bigr|i\Bigr\rangle\Biggr|^2
     \delta(E_i-E_f),
   \label{E:W}
\end{equation}
where $i$ and $f$ denote the initial and final states of the
unperturbed electronic-nuclear spin system.
Then we obtain $T_1=(I-m)(I+m+1)/2W$.
Considering the significant difference between the nuclear and electronic
energy scales at moderate fields and assuming the Fourier components of
the coupling tensors to have little momentum dependence as
$\sum_n e^{{\rm i}k(n+i/2-1)}A_{n:i}^\lambda
 \equiv A_{k,i}^\lambda\simeq A_i^\lambda$,
the Raman and three-magnon relaxation rates read
\begin{eqnarray}
   &&
   \frac{1}{T_1^{(2)}}
   \simeq\frac{2(g\mu_{\rm B}\hbar\gamma_{\rm N})^2}{\hbar N}
    \sum_{k_1}\sum_{\sigma=\pm}\sum_{i=1,3}
     |W_{ii}(k_1,\sigma k_2^{(i)})|^2
   \nonumber\\
   &&\qquad\times
     (\bar{n}_{k_1:i}+1)\bar{n}_{k_2^{(i)}:i}
     \left|
      \frac{{\rm d}\omega_{k:i}}{{\rm d}k}
     \right|_{k=k_2^{(i)}}^{-1},
   \label{E:T1(2)}
   \\
   &&
   \frac{1}{T_1^{(3)}}
   \simeq\frac{(g\mu_{\rm B}\hbar\gamma_{\rm N})^2}{16\hbar SN^2}
    \sum_{k_1,k_2}\sum_{\sigma=\pm}
    \Biggl[
     2|W_{111}(k_1,k_2,\sigma k_3^{(1)})|^2
   \nonumber\\
   &&\qquad\times
     \bar{n}_{k_1:1}(\bar{n}_{k_2:1}+1)(\bar{n}_{k_3^{(1)}:1}+1)
     \left|
      \frac{{\rm d}\omega_{k:1}}{{\rm d}k}
     \right|_{k=k_3^{(1)}}^{-1}
   \nonumber\\
   &&\qquad
    + |W_{132}(k_1,k_2,\sigma k_3^{(2)})|^2
     \bar{n}_{k_1:1}\bar{n}_{k_2:3}(\bar{n}_{k_3^{(2)}:2}+1)
   \nonumber\\
   &&\qquad\times
     \left|
      \frac{{\rm d}\omega_{k:2}}{{\rm d}k}
     \right|_{k=k_3^{(2)}}^{-1}
    + |W_{313}(k_1,k_2,\sigma k_3^{(3)})|^2
   \nonumber\\
   &&\qquad\times
     (\bar{n}_{k_1:3}+1)(\bar{n}_{k_2:1}+1)\bar{n}_{k_3^{(3)}:3}
     \left|
      \frac{{\rm d}\omega_{k:3}}{{\rm d}k}
     \right|_{k=k_3^{(3)}}^{-1}
    \Biggr],
   \label{E:T1(3)}
\end{eqnarray}
where $\bar{n}_{k:i}\equiv\langle\alpha_{k:i}^\dagger\alpha_{k:i}\rangle$
is the thermal distribution function of modified spin waves, while
$k_2^{(i)}$ and $k_3^{(i)}$ are determined through
$\omega_{k_1:i}-\omega_{k_2^{(i)}:i}-\hbar\omega_{\rm N}=0$,
$\omega_{k_3^{(1)}:1}+\omega_{k_2:1}-\omega_{k_1:1}
-\hbar\omega_{\rm N}=0$,
$\omega_{k_3^{(2)}:2}-\omega_{k_2:3}-\omega_{k_1:1}
-\hbar\omega_{\rm N}=0$, and
$\omega_{k_3^{(3)}:3}-\omega_{k_2:1}-\omega_{k_1:3}
+\hbar\omega_{\rm N}=0$.
$W_{ii}(k_1,k_2)$ and $W_{ijl}(k_1,k_2,k_3)$ are diagrammatically
represented in Fig. \ref{F:diagram}.
Within the first-order mechanism containing a nuclear spin in direct
contact with spin waves via the hyperfine interaction, any multi-magnon
relaxation rate is much smaller than the Raman one.
However, some of multi-magnon processes make a significant contribution to
$1/T_1$ through the second-order mechanism, where a nuclear spin flips
with the help of {\it virtual spin waves} which are then scattered
thermally via the four-magnon exchange interaction.
We consider the leading second-order relaxation, that is, 
{\it exchange-scattering-induced three-magnon processes}, as well as the
first-order relaxation.
Second-order single-magnon and Raman relaxation processes, containing
three and two virtual magnons, respectively, are much more accidental due
to the momentum conservation and much less contributive due to the magnon
series damping.
As for four-magnon processes, the first-order relaxation is nonexistent to
begin with, whereas the second-order one originates in the six-magnon
exchange interaction and contains two virtual magnons.
Thus and thus, all other higher-order processes have no significant effect
on the relaxation scenario.
\begin{figure}
\centering
\includegraphics[width=84mm]{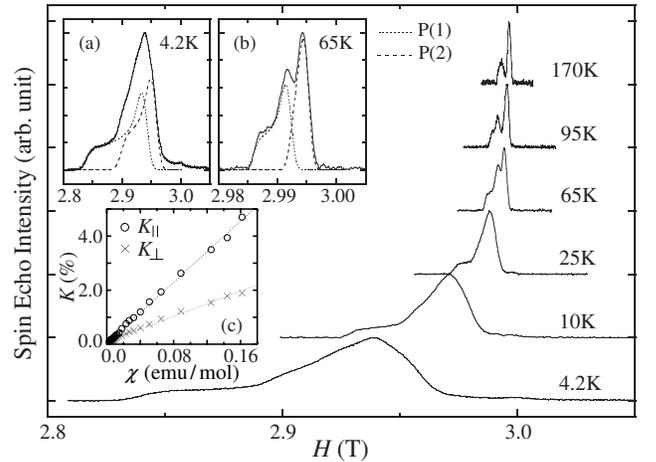}
\vspace*{-3mm}
\caption{$^{31}\mbox{P}$ NMR spectra measured at $51.711\,\mbox{MHz}$.
         They can be decomposed into typical powder patterns [the insets
         (a) and (b)], originating in the P(1) (dotted lines) and P(2)
         (broken lines) sites, and the NMR shift for the latter is plotted
         as a function of the susceptibility at $H=3\,\mbox{T}$
         [the inset (c)].}
\label{F:spectra}
\end{figure}

\section{Experimental Test}

   Now we present $^{31}\mbox{P}$ NMR spectra for powder samples of
Ca$_3$Cu$_3$(PO$_4$)$_4$ in Fig. \ref{F:spectra}.
The crystallographically inequivalent P sites, labeled P(1) and P(2), give
two distinct lines, each of which has a characteristic shape of the
anisotropic powder pattern, as is revealed in the insets (a) and (b).
Both lines broaden and shift to lower field with decreasing temperature,
implying that contact as well as dipolar terms exist in the hyperfine
field on the P nuclei.
The NMR shifts are thus anisotropic and their parallel ($K_\parallel$) and
perpendicular ($K_\perp$) components for the P(2) sites are plotted as
functions of the susceptibility at $H=3\,\mbox{T}$ in the inset (c).
If we roughly estimate the hyperfine coupling between each P(2) nucleus
and the nearest-neighbor Cu$^{\rm II}$ ion through the relations
$A_\perp=N_{\rm A}\mu_{\rm B}{\rm d}K_\perp/{\rm d}\chi$ and
$A_\parallel=N_{\rm A}\mu_{\rm B}{\rm d}K_\parallel/{\rm d}\chi$,
where $N_{\rm A}$ is the Avogadro number, we find that
$A_\perp=0.66\,\mbox{kOe}/\mu_{\rm B}$ and
$A_\parallel=1.58\,\mbox{kOe}/\mu_{\rm B}$.
In order to reveal the low-frequency spin dynamics, we measure $T_1$ for
the P(2) sites through the saturation-recovery method.
\begin{figure}
\centering
\includegraphics[width=86mm]{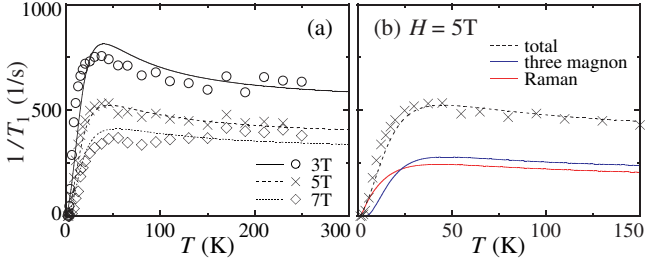}
\vspace*{-3mm}
\caption{(Color online)
         Experimental (symbols) and theoretical (lines) findings of
         $1/T_1$ as functions of temperature.
         $1/T_1^{(2)}$ and $1/T_1^{(3)}$ are also plotted in (b).}
\label{F:T1T}
\end{figure}
\begin{figure}
\centering
\includegraphics[width=86mm]{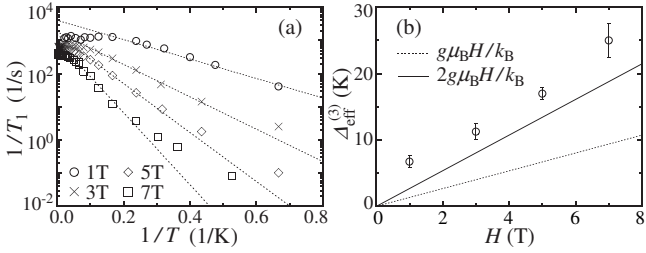}
\vspace*{-3mm}
\caption{(a) Semilog plots of $1/T_1$ as functions of $1/T$.
             Dotted lines are guides to the probable exponential behavior
             $1/T_1\propto{\rm e}^{-{\mit\Delta}_{\rm eff}^{(3)}/T}$ at
             each field.
         (b) The thus-extracted ${\mit\Delta}_{\rm eff}^{(3)}$ as a
             function of field.
             The two slopes,
             ${\mit\Delta}_{\rm eff}^{(3)}= g\mu_{\rm B}H/k_{\rm B}$ and
             ${\mit\Delta}_{\rm eff}^{(3)}=2g\mu_{\rm B}H/k_{\rm B}$,
             are shown for reference.}
\label{F:T1gap}
\end{figure}

   Temperature dependences of $1/T_1$ are shown in Fig. \ref{F:T1T}.
In each Cu$^{\rm II}$ trimer unit, the two spins
$\mbox{\boldmath$S$}_{n:1}$ and $\mbox{\boldmath$S$}_{n:2}$
are almost equidistant from the nearby P(2) atom, whereas the other one
$\mbox{\boldmath$S$}_{n:3}$ is much more distant from that. \cite{A29}
Therefore, the isotropic coupling constants are taken as
$A_1^-=A_2^-\equiv A^-$ and $A_3^-=0$.
Since dipolar interactions are also sensitive to the location of
correlating moments, the anisotropic coupling constants may be taken
similarly as $A_1^z=A_2^z\equiv A^z$ and $A_3^z=0$.
Then we set $A^-$ and $A^z$ equal to $0.85\,\mbox{kOe}/\mu_{\rm B}$ and
$1.20\,\mbox{kOe}/\mu_{\rm B}$, respectively, which are both consistent
well with the experimental findings $A_\perp=0.66\,\mbox{kOe}/\mu_{\rm B}$
and $A_\parallel=1.58\,\mbox{kOe}/\mu_{\rm B}$.
Considering that recent electron-spin-resonance measurements of this
compound \cite{O} have yield temperature-dependent and anisotropic $g$
values ($g_\parallel>g_\perp\simeq 2$), the $A_\parallel/A_\perp$ value
may be closer to the theoretical parametrization.
We are further convinced of the coupling constants employed finding the
nearest Cu(1)-P(2) distance to be about $(1/A^z)^{1/3}=2.5\,\mbox{\AA}$,
which is in excellent agreement with the crystalline structure. \cite{A29}
The thus-calculated $1/T_1=1/T_1^{(2)}+1/T_1^{(3)}$ reproduces the
observations pretty well.
The exchange-scattering-enhanced three-magnon relaxation rate generally
grows into a major contribution to $1/T_1$ with increasing temperature
and decreasing field.
While both $1/T_1^{(2)}$ and $1/T_1^{(3)}$ exhibit an exponential behavior
at low temperatures, their activation energies, referred to as
$k_{\rm B}{\mit\Delta}_{\rm eff}^{(2)}$ and
$k_{\rm B}{\mit\Delta}_{\rm eff}^{(3)}$, respectively, look different.
At moderately low temperatures and weak fields,
$\hbar\omega_{\rm N}\ll k_{\rm B}T\ll J_2$, Eq. (\ref{E:T1(2)}) reads
\begin{equation}
   \frac{1}{T_1^{(2)}}\simeq
   \frac{J_1+J_2}{2\pi\hbar SJ_1J_2}
   (g\mu_{\rm B}\hbar\gamma_{\rm N}A^z)^2
   {\rm e}^{-g\mu_{\rm B}H/k_{\rm B}T}
   K_0\Bigl(\frac{\hbar\omega_{\rm N}}{2k_{\rm B}T}\Bigr),
   \label{E:T1(2)ap}
\end{equation}
where $K_0$ is the modified Bessel function of the second kind and behaves
as
$K_0(\hbar\omega_{\rm N}/2k_{\rm B}T)
 \simeq 0.80908-{\rm ln}(\hbar\omega_{\rm N}/k_{\rm B}T)$.
Thus we learn that
${\mit\Delta}_{\rm eff}^{(2)}\simeq g\mu_{\rm B}H/k_{\rm B}$.
Equation (\ref{E:T1(3)}) is much less analyzable, but Fig. \ref{F:T1T}(b)
claims that ${\mit\Delta}_{\rm eff}^{(2)}<{\mit\Delta}_{\rm eff}^{(3)}$.
Figure \ref{F:T1gap} brings the next leading exponential behavior
$1/T_1\propto{\rm e}^{-{\mit\Delta}_{\rm eff}^{(3)}/T}$ to light, because
Eq. (\ref{E:T1(2)ap}) is valid for $T\ll J_2/k_{\rm B}=8\,\mbox{K}$.
${\mit\Delta}_{\rm eff}^{(3)}$ looks like $2g\mu_{\rm B}H/k_{\rm B}$ or
more, rather than $g\mu_{\rm B}H/k_{\rm B}$.
The two-magnon-mediated nuclear spin relaxation is mainly given by
$W_{11}(k_1,k_2)$, where
a ferromagnetic spin wave of energy $\omega_{k_1:1}$ is created and
that of energy $\omega_{k_2:1}\simeq\omega_{k_1:1}$ is destructed,
while the three-magnon-mediated one by $W_{111}(k_1,k_2,k_3)$, where
two net spin waves of energy $\omega_{k_2:1}$ and $\omega_{k_3:1}$ are
created and that of energy
$\omega_{k_1:1}\simeq\omega_{k_2:1}+\omega_{k_3:1}$ is destructed.
Therefore, Raman processes are activated by the energy
$\omega_{k=0:1}=g\mu_{\rm B}H$, which is consistent with
Eq. (\ref{E:T1(2)ap}), whereas three-magnon ones roughly by the energy
$2\omega_{k=0:1}=2g\mu_{\rm B}H$, which may look somewhat larger due to
the complicated field dependence lying in the second-order mechanism.
\begin{figure}
\centering
\includegraphics[width=86mm]{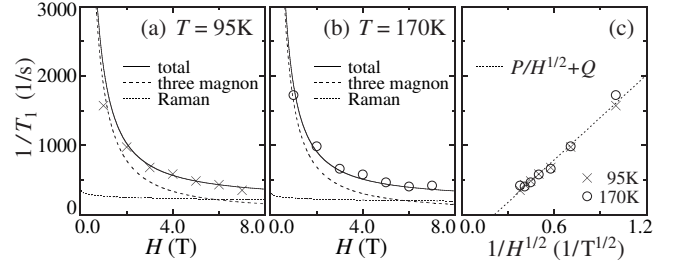}
\vspace*{-3mm}
\caption{Experimental (symbols) and theoretical (lines) findings of
         $1/T_1$ as functions of field at $95\,\mbox{K}$ (a) and
         $170\,\mbox{K}$ (b), where $1/T_1^{(2)}$ and $1/T_1^{(3)}$ are
         also plotted.
         The observations are replotted as functions of $1/\sqrt{H}$ and
         fitted to an expression $1/T_1=P/\sqrt{H}+Q$ (c).}
\label{F:T1H}
\end{figure}

   Field dependences of $1/T_1$ at higher temperatures more impress on us
the significance of three-magnon processes.
Figures \ref{F:T1H}(a) and \ref{F:T1H}(b) show that the accelerated
relaxation with decreasing field can never be explained by
the Raman scheme but should be attributed to exchange-scattering-enhanced
three-magnon processes.
The spin-diffusion model may be mentioned in this context.
Diffusion-dominated 1D spin dynamics gives $1/T_1$ of the form
$P/\sqrt{H}+Q$, \cite{H965} where the first and second terms come from
transverse and longitudinal spin fluctuations, respectively, and are both
positive.
Figure \ref{F:T1H}(c) shows that the present observations fitted to the
diffusive law result in negative $Q$.
We do not exclude a possibility of diffusive dynamics appearing in 1D
ferrimagnets as well, but a distinct field dependence of the second-order
relaxation mechanism masks such a moderate field effect in the present
case.
It was in AgVP$_2$S$_6$ rather than in the most familiar Haldane-gap
antiferromagnet Ni(C$_2$H$_8$N$_2$)$_2$NO$_2$ClO$_4$, whose excitation
spectrum drastically varies with increasing field, that integral-spin
diffusive correlations were observed. \cite{T2173}
Exchange-scattering-induced three-magnon processes are sensitive to an
excitation gap and their contribution to $1/T_1$ is strongly suppressed,
for example, by slight magnetic anisotropy.
There are indeed some indications of spin diffusion \cite{F8410}
in the ferrimagnetic chain compound
Mn(C$_5$H$_2$O$_2$F$_6$)$_2$C$_{10}$H$_{17}$N$_2$O$_2$ with nonnegligible
single-ion anisotropy. \cite{C1756}

\section{Concluding Remarks}

   We have performed NMR measurements on the topological ferrimagnet
Ca$_3$Cu$_3$(PO$_4$)$_4$ and have confirmed a novel scenario for 1D spin
dynamics$-${\it multimagnon-mediated nuclear spin relaxation}, by showing
the parametrization to be crystallographically convincing, revealing that
the relaxation is activated by twice the gap rather than the gap itself at
low temperatures, while it is remarkably accelerated with decreasing field,
and pointing out the irrelevance of the spin-diffusion model.
Indeed pioneering $T_1$ observations on the layered ferromagnet
CrCl$_3$ (Ref. \onlinecite{N354}) and the coupled-chain antiferromagnet
CsMnCl$_3\cdot$2H$_2$O (Ref. \onlinecite{N5325}) claimed to have detected
three-magnon processes, but they were both, in some sense, classical
findings under the 3D long-range order. \cite{P398}
Our findings are literally 1D quantum spin relaxation beyond the Raman
mechanism, which were obtained through an elaborately modified spin-wave
theory. \cite{Y064426}

   The bond-alternating homometallic chain compound
Cu(C$_5$ClH$_4$N)$_2$(N$_3$)$_2$ (Ref. \onlinecite{E4466}) is another
anisotropy-free ferrimagnet of topological origin \cite{N214418} and
therefore NMR measurements on it are highly encouraged.
Since 1D ferromagnets, which can be regarded as low-energy sectors of 1D
ferrimagnets, may also play this fascinating scenario, more understanding
will come with further experiments, for instance, on the
spin-$\frac{1}{2}$ ferromagnetic chain compound
(CH$_3$)$_4$NCuCl$_3$. \cite{L463}
The quasi-one-dimensional mixed-spin ferromagnet
$\mbox{MnNi(NO}_2\mbox{)}_4\mbox{(C}_2\mbox{H}_8\mbox{N}_2\mbox{)}_2$
(Ref. \onlinecite{K1530}) is also highly interesting in this context,
whose low-energy spectrum consists of two dispersive ferromagnetic
excitation branches. \cite{F174430,Y054423}
\acknowledgments

   This work was supported by the Ministry of Education, Culture, Sports,
Science, and Technology of Japan.

\end{document}